\documentclass[lettersize,journal]{IEEEtran}
\usepackage{amsmath,amsfonts}
\usepackage{algorithm}
\usepackage{array}
\usepackage[caption=false,font=normalsize,labelfont=sf,textfont=sf]{subfig}
\usepackage{textcomp}
\usepackage{stfloats}
\usepackage{url}
\usepackage{verbatim}
\usepackage{graphicx}
\usepackage{cite}
\usepackage{amsthm}
\usepackage{amsmath}
\usepackage{autobreak}
\usepackage{amsfonts} 
\usepackage{amssymb} 
\allowdisplaybreaks
\usepackage{adjustbox}
\usepackage{booktabs}
\usepackage{makecell}
\usepackage{siunitx}
\usepackage{multirow}
\usepackage{balance}
\usepackage{longtable}

\usepackage{algorithm}
\usepackage{algorithmicx}
\usepackage{algpseudocode}
\usepackage{bm}
\usepackage{graphicx}
\usepackage{placeins}
\usepackage{amsmath}

\usepackage{booktabs}
\usepackage{tabularx}
\usepackage{listings}
\usepackage{hyperref}
\usepackage{wasysym}
\usepackage{array}
\usepackage{tabularx}
\usepackage{caption}
\usepackage{xcolor}
\usepackage{amsthm}

\hypersetup{
  colorlinks,
  linkcolor={blue!70!green},
  citecolor={green!70!blue},
  urlcolor={orange!70!red}
}

\newcommand{\para}[1]{\vspace{4pt}\noindent\textbf{{#1}}}
\newcommand{\ignore}[1]{}

\usepackage{amsmath,amsfonts,bm}

\def\eqref#1{equation~\ref{#1}}

\def\1{\bm{1}}

\DeclareMathAlphabet{\mathsfit}{\encodingdefault}{\sfdefault}{m}{sl}
\SetMathAlphabet{\mathsfit}{bold}{\encodingdefault}{\sfdefault}{bx}{n}

\begin{document}
\bstctlcite{IEEEexample:BSTcontrol}
\title{Training-Data Extraction from Genomic Language Models}
\author{
\IEEEauthorblockN{
Tianhao Mao\textsuperscript{1*},
Zihao Wang\textsuperscript{2*},
Dongfang Zhao\textsuperscript{1},
Rui Zhu\textsuperscript{3},
Xiaofeng Wang\textsuperscript{2},
Haixu Tang\textsuperscript{1},
}

\IEEEauthorblockA{
\textsuperscript{1}Luddy School of Informatics, Computing, and Engineering, Indiana University Bloomington\\
\textsuperscript{2}College of Computing and Data Science, Nanyang Technological University\\
\textsuperscript{3}Yale School of Medicine, Yale University\\
Email: tianmao@iu.edu, zihao.wang@ntu.edu.sg, zhaodo@iu.edu, rui.zhu.rz399@yale.edu, xiaofeng.wang@ntu.edu.sg, hatang@iu.edu}
\thanks{\textsuperscript{*}The first two authors are co-first authors contributed equally to this work.}
}
\maketitle

\begin{abstract}
Genomic language models generate long nucleotide continuations from short prefixes, exposing a direct route for studying training-sequence extraction through the generation interface. We present an output-only black-box attack on Evo 2 that selects convergent groups of repeated generations without gradients, token likelihoods, or training-corpus access during candidate selection. Recovered content is then scored by an evaluator combining aligned length, nucleotide identity, match multiplicity, and sequence complexity. Across 304 runs on Evo 2 7B with bacteriophage and plasmid prefixes, the attack yields 888 candidate clusters, of which 85 match training segments of at least 500 bp, giving a cluster-level attack success rate of $9.57\%$. Final filtering retains 22 uniquely matched extraction events at the 500-bp threshold and 15 at 1,000 bp, covering 14 and 7 distinct training-sequence regions, respectively. Within-category source-homology screening retains 11 of the 22 events at a $99\%$ exclusion threshold and four at $98\%$, with the latter mapping to two reference regions. The results establish convergence among repeated nucleotide generations as a practical output-only signal for recovering long training-sequence segments and characterize how recovery depends on prefix length, cluster size, perplexity, early stopping, and sequence domains.
\end{abstract}

\section{Introduction}
\label{sec:introduction}

Genomic language models learn statistical dependencies in nucleotide sequences from large biological corpora, supporting tasks ranging from functional annotation to sequence design~\cite{bioinformatics2021-1,naturemethod2025,science2024}. Autoregressive models such as Evo 2 go further, generating long DNA continuations from short nucleotide prefixes~\cite{nature2026-1}. This generative capabilities also raise a data-confidentiality question: can an adversary recover substantial portions of training sequences through access to a model's generation interface?

We select Evo 2 as the target of this study for three reasons. First, it generates autoregressively at single-nucleotide resolution, producing exactly the long continuations from short prefixes that our extraction setting requires. Comparable models do not expose this interface: DNABERT and Nucleotide Transformer primarily learn masked sequence representations, and AlphaGenome predicts functional properties from a given input sequence~\cite{bioinformatics2021-1,naturemethod2025,nature2026-2}. Second, while Evo was trained on prokaryotic and bacteriophage sequences, Evo 2 draws on data spanning all domains of life, letting us examine recovery across diverse sequence collections within a single model~\cite{science2024,nature2026-1}. Third, the released model parameters, inference code, and OpenGenome2 training corpus make generation experiments reproducible and allow recovered content to be checked directly against the training data~\cite{nature2026-1,arc2025opengenome2}. We use the 7B version so that extensive repeated generation experiments computationally practical. These resources support our experimental evaluation: the attack itself selects candidates from generated nucleotide sequences alone.

Training-data extraction has been demonstrated in both language and image generation. Prior attacks on language models generate candidate outputs and ranking them using memorization signals~\cite{usenix2021}, attacks on diffusion models exploit similarity among repeated generations to identify recoverable training images~\cite{usenix2023}. Neither setting has been examined for genomic models. Security studies of such models have instead addressed jailbreaks and the robustness of generation safeguards~\cite{ICLR2026,NIPSworkshop2025}, leaving open a distinct question: whether a fixed pretrained model can disclose nucleotide content from its training corpus. The question also differs from membership inference, which asks only whether a supplied record was used in training~\cite{oakland2017,oakland2022}. An extraction attack must produce the sequence itself. 

Nucleotide sequences complicate this task. Repeated generations can agree closely for reasons unrelated to memorization: short repetitive regions and low-complexity regions occur throughout genomes, and related records share long homologous segments. High similarity is therefore weak evidence on its own, and recovered content requires validation beyond raw sequence agreement. A procedure suited to this setting must consider both nucleotide identity and the extent of the aligned region, tolerate insertions and deletions, and account for sequence complexity and match multiplicity. Moreover, it must also operate on nucleotide outputs alone, since an adversary querying a generation service observes no token likelihoods or gradients.

In this work, we investigate training-data extraction from Evo 2 through a cluster-based black-box attack. The adversary constructs short prefixes from an auxiliary sequence collection and repeatedly samples continuations for each prefix. The central idea is to use convergence across repeated generations as an observable signal for candidate selection. We build a similarity graph over the continuations and retain clusters whose members satisfy pairwise nucleotide-identity and alignment-coverage requirements. Candidate selection uses the generated sequences alone. For evaluation, we match the selected continuations against a separate subset of the model's training data and apply criteria based on aligned length, match multiplicity, and sequence complexity. Input prefixes are excluded from the analyzed outputs, so the measured extraction concerns nucleotide content generated beyond the prompt.

We evaluate the attack on the Evo 2 7B model using bacteriophage and plasmid training sequences. Across 304 runs with 300-bp and 500-bp prefixes, the default configuration yields 888 candidate clusters. Of these, 85 match training segments of at least 500 bp under the 99\% identity criterion, a cluster-level success rate of 9.57\%; 45 match segments of at least 1,000 bp, and the longest extracted spans reach 5,000 bp, the full continuation length. We then apply a stricter unique-match criterion, requiring that a continuation meet the length threshold, match a single record in the searched validation subset, and pass all final filters. This yields 22 extraction events at the 500-bp threshold, of which 15 also hold at 1,000 bp, and merging overlapping recoveries within each training record leaves 14 and 7 distinct regions, respectively. To test whether these recoveries reflect memorization rather than homology, we exclude near-identical sequences from the corresponding source collection within the same category: 11 of the 22 events survives at a $99\%$ exclusion threshold and four at $98\%$.

We further characterize how prefix length and selection stringency shape extraction outcomes. Extending prefixes from 300 to 500 bp yields more extractions reaching 500 bp, though most of the gain falls below 1,000 bp. Requiring larger clusters concentrates successful extractions in a smaller candidate set while retaining fewer of them overall. An auxiliary experiment with greedy generation shows ranking continuations by lowest-perplexity predominantly favors repetitive sequences, which is what motivates our explicit complexity filtering. We also examine early stopping during generation and apply the extraction procedure to mRNA sequences.

Our contributions are threefold:
\begin{itemize}
    \item We formulate training sequence extraction from genomic language models as an output-only black-box attack on training data confidentiality, in which the adversary supplies short nucleotide prefixes and observes generated continuations with no access to token likelihoods or model internals.
    \item We develop a convergence-based candidate-selection procedure tailored to nucleotide sequences, combining pairwise identity and alignment coverage to tolerate mismatches and indels, together with a validation framework that quantifies aligned length, sequence complexity, match multiplicity, and repeated recovery of the same region.
    \item We evaluate Evo 2 across 304 runs on bacteriophage and plasmid data, characterize how extraction varis across prefix length, cluster size, perplexity, and early stopping, and extend the evaluation to mRNA sequences.
\end{itemize}


\section{Background and Related Work}
\label{sec:background}

\subsection{Genomic Language Models}
Genomic language models learn statistical dependencies in nucleotide sequences through self-supervised pretraining. These models support tasks such as identifying regulatory elements, predicting variant effects, and generating DNA sequences. DNABERT~\cite{bioinformatics2021-1} represents DNA using overlapping $k$-mers and applies masked language modeling to learn representations that can be adapted to downstream genomic prediction tasks. Nucleotide Transformer~\cite{naturemethod2025} extends this approach through a family of models pretrained on large human and multispecies genome collections, demonstrating the utility of pretrained sequence representations across diverse functional prediction tasks.

Sequence-to-function models provide a complementary direction. AlphaGenome predicts expression, splicing, chromatin, and other regulatory signals from megabase-scale DNA inputs for variant-effect analysis~\cite{nature2026-2}. Because its output is a set of functional predictions rather than an autoregressive nucleotide continuation, it represents a different privacy and attack surface from the generative model considered here.

Autoregressive genomic models additionally support sequence generation by predicting successive nucleotides conditioned on preceding context. Evo~\cite{science2024} established this approach at genomic scale, using prokaryotic and bacteriophage sequences to learn relationships spanning molecular functions and genome organization. Evo 2~\cite{nature2026-1} extends genomic modeling across all domains of life. Its 7B and 40B models use the StripedHyena~2 architecture, operate at single-nucleotide resolution, and support context lengths of up to one million nucleotides. This generative interface allows an adversary to supply a short DNA prefix and obtain a substantially longer continuation, making training-sequence recovery a relevant security question.

Evo 2 is trained on OpenGenome2, a corpus containing more than 8.8 trillion nucleotides assembled from public biological resources~\cite{nature2026-1,arc2025opengenome2}. Its sources include prokaryotic genomes from the Genome Taxonomy Database (GTDB), metagenomic sequences, and eukaryotic sequences from NCBI and Ensembl. OpenGenome2 also incorporates the earlier OpenGenome collection, including curated bacteriophage sequences from IMG/VR and plasmid sequences from IMG/PR. These phage and plasmid components are distributed as part of the released Evo 2 training corpus~\cite{arc2025opengenome2}. Our evaluation focuses on extracting bacteriophage and plasmid training sequences from Evo 2.

Recent genomic-model security studies have examined a different threat class. GeneBreaker evaluates jailbreaks that guide DNA language models toward pathogen-related outputs~\cite{ICLR2026}, while Black et al.~\cite{NIPSworkshop2025} study the robustness of data-exclusion safeguards under adversarial fine-tuning of open-weight genome models. These studies concern safety-policy evasion or changes to model parameters; they do not test whether a fixed pretrained model reproduces sequences from its training corpus.

\subsection{Training-Data Extraction Attacks}
Training-data extraction aims to recover training content through access to a trained model. For generative models, an attacker can elicit outputs and identify those that reproduce individual training examples or substantial portions of them. Carlini et al.~\cite{usenix2019} studied unintended memorization by inserting rare canary sequences into training data and measuring their exposure through their likelihood ranking among possible secrets. Their experiments showed that memorized sequences could be recovered from neural sequence models, establishing a connection between training-data memorization and content disclosure.

Subsequent work demonstrated extraction from large language models trained on naturally occurring text. Carlini et al.~\cite{usenix2021} recovered hundreds of verbatim training examples from GPT-2 using a generate-and-filter procedure. They sampled candidate outputs and prioritized likely memorized text using scores based on model likelihoods, comparisons with other models, and compression. A further study~\cite{ICLR2023} systematically examined how recoverable memorization varies with model size, training-data duplication, and the amount of conditioning context. These findings motivate examining prefix length in genomic extraction, where a relatively short observed sequence may elicit a much longer continuation matching training content.

Extraction has also been demonstrated for image diffusion models. Carlini et al.~\cite{usenix2023} used repeated generation and similarity-based filtering to identify memorized images, with groups of nearly identical outputs serving as a signal of reproducible training content. Our approach adapts this general principle to nucleotide continuations: continuations from the same prefix are compared through sequence alignment, and clusters satisfying identity and alignment-coverage requirements are selected as extraction candidates. The method uses generated sequences without requiring token likelihoods. In genomic data, biological homology and repetitive sequence can also produce high similarity, requiring explicit assessment of complexity and match multiplicity. Our candidate-selection and validation stages address these sequence-specific issues through local alignments, indel-aware identity, complexity filters, and matched-record counts, as detailed in Section~\ref{sec:method}.

Nemecek et al.~\cite{BCB2026} directly study memorization and privacy risks in genomic language models through a framework combining perplexity-based detection, canary extraction, and membership inference. They evaluate SimpleDNALM, DNABERT-2, HyenaDNA, and Evo in controlled training and fine-tuning experiments, using LoRA for Evo. Their protocol inserts 64-nucleotide random canaries at different repetition rates into small genomic training corpora. Canary extraction conditions on a prefix of the planted sequence and uses beam search and model-likelihood ranking to assess recovery. The framework aggregates the three evaluation vectors into a maximum vulnerability score, characterizing how duplication, architecture, and adaptation affect memorization risk.

Our study addresses extraction of naturally occurring training sequences from the released Evo 2 7B model with its pretrained parameters unchanged. We construct prompts from an auxiliary genomic collection and discover candidate training segments through alignment-based convergence among repeated continuations, using generated nucleotide sequences alone for candidate selection. The evaluator then verifies recovered content against a separate training-data subset using alignment length, identity, match multiplicity, and sequence complexity. This complements controlled canary auditing with an output-only extraction attack that recovers long genomic segments, reaching 5,000 bp in the bacteriophage and plasmid evaluation, and is additionally evaluated on mRNA sequences.

\subsection{Membership Inference Attacks}
Membership inference attacks (MIAs) determine whether a given candidate record was used to train a model. Shokri et al.~\cite{oakland2017} introduced a black-box approach that trains shadow models to reproduce the target model's behavior and uses their prediction outputs to learn a membership classifier. Carlini et al.~\cite{oakland2022} subsequently proposed the likelihood-ratio attack, which compares the target model's score for a candidate with score distributions estimated under membership and nonmembership hypotheses. Duan et al.~\cite{COLM2024} found that common MIA methods were close to chance in many language-model pretraining settings and showed that distribution shifts between member and nonmember samples can inflate apparent attack performance. Such attacks assess training participation for a supplied record rather than recovering the record itself.

Genomic privacy work has also considered membership without a generative model. Bu et al.~\cite{bioinformatics2021-2} infer haplotype membership from allele-frequency summaries or Beacon responses and reconstruct haplotypes from aggregate genomic statistics. At the level of deployed medical models, Knolle et al.~\cite{nature2026-3} show that aggregate MIA performance can conceal substantial patient-level variation in privacy risk. These studies emphasize membership or re-identification of supplied individuals and records, whereas the present task recovers nucleotide content from model generations.

Membership inference and extraction are related because membership scores can help rank generated candidates, as in prior language-model extraction attacks~\cite{usenix2021}. Their objectives and outputs nevertheless differ: model-membership inference starts with a supplied candidate record and returns a membership decision or score, whereas extraction seeks to recover training content through model outputs. In our setting, the adversary supplies an auxiliary genomic prefix and obtains nucleotide continuations containing training segments beyond the prompt. The attack selects candidate continuations through convergence among repeated generations. Matching against the training-data validation subset is performed subsequently by the evaluator to measure the recovered content; it is not a membership-query interface available to the adversary. The attack thus outputs candidate extracted sequences whose matches to training data are assessed separately by the evaluator.

%

\section{Method}
\label{sec:method}
\subsection{Threat Model}
We consider an adversary with output-only black-box access to a pretrained genomic language model. The adversary can submit nucleotide prefixes and observe the resulting continuations, and constructs these prefixes from public sequences or sequences already in their possession. The adversary has no direct access to the training corpus or training membership labels. The attack relies solely on generated sequences and does not require model parameters, gradients, or token likelihoods.

Let $f_\theta$ denote the target model, $D_{\mathrm{train}}$ its training corpus, and $D_{\mathrm{src}}$ the attacker's source collection. Given a prefix $x$ sampled from a sequence in $D_{\mathrm{src}}$, the adversary seeks a continuation $y\sim f_\theta(\cdot\mid x)$ containing a long segment of a training sequence that was not supplied in the prompt. The attack outputs candidate continuations that may contain memorized training-sequence content; their matches to training data are assessed subsequently by the evaluator. Repeated generation of highly similar continuations provides the signal used to identify these candidates.

For evaluation, we partition a subset of the model's training data into the source collection $D_{\mathrm{src}}$ and a validation collection $D_{\mathrm{val}}$. This partition constructs the evaluation benchmark; $D_{\mathrm{val}}$ is never used for candidate generation or selection. Thus, $D_{\mathrm{val}}\subseteq D_{\mathrm{train}}$ and $D_{\mathrm{src}}\cap D_{\mathrm{val}}=\emptyset$ at the sequence-record level. The adversary uses only $D_{\mathrm{src}}$ to construct prompts and is not supplied its training membership labels. After generation and candidate selection, the evaluator matches the candidates against $D_{\mathrm{val}}$ to determine extraction success according to the criteria defined below. 

\subsection{Overall Pipeline}
\begin{figure*}[t]
\centering
\includegraphics[width=0.97\textwidth]{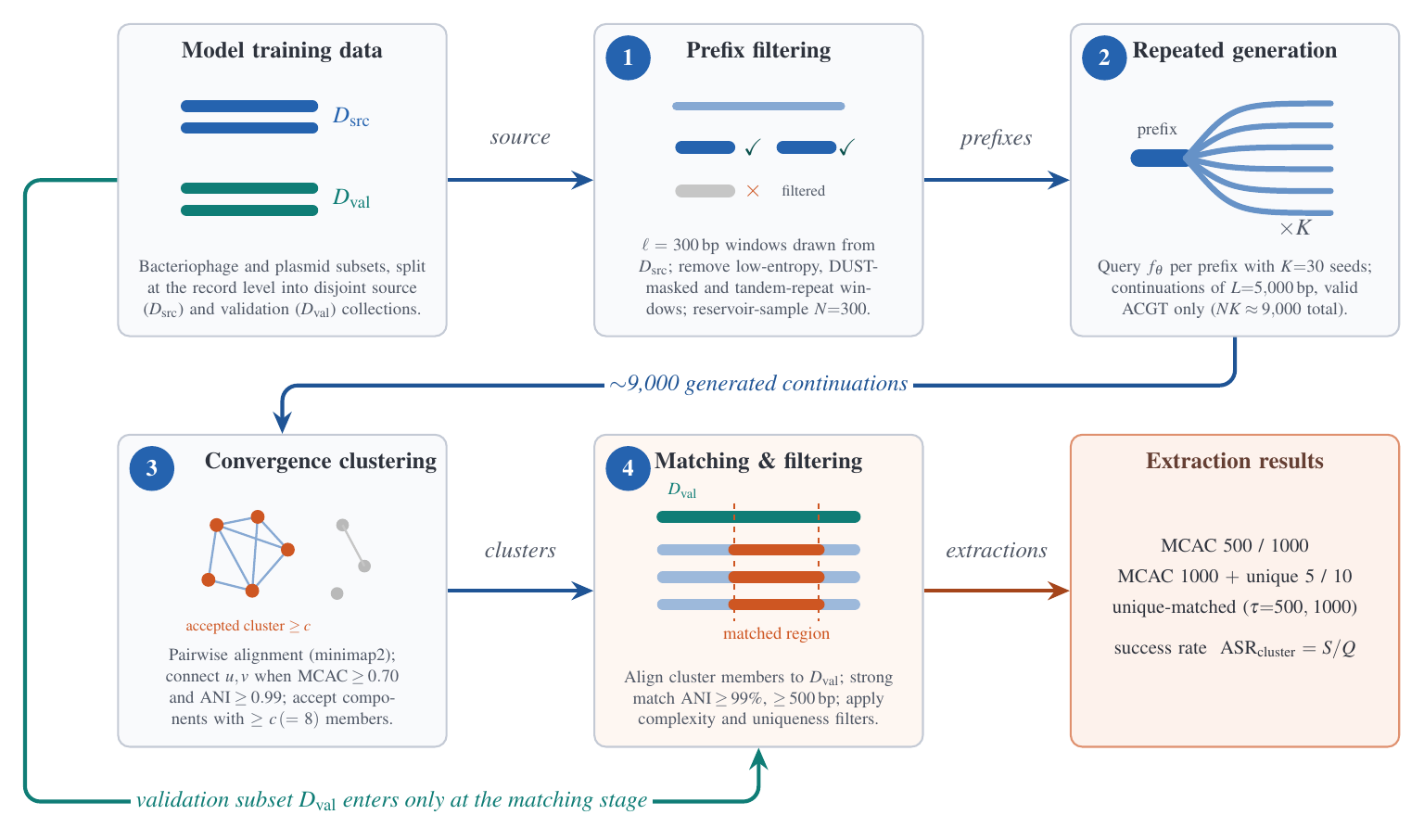}
\caption{Overview of the cluster-based training-data extraction pipeline. Candidate generation and selection use only source prefixes and model outputs. An accepted component must contain at least $c=8$ members in the default setting, and \emph{every pair} must pass both identity and coverage thresholds. Training-data validation and final extraction filtering are performed by the evaluator.}
\label{fig:pipeline}
\end{figure*}

The pipeline consists of prefix construction, repeated generation, convergence clustering, and training-data validation, as illustrated in Figure~\ref{fig:pipeline}. Generation and clustering use only the source prefixes and model outputs. The evaluator subsequently uses the validation collection to measure recovery of training-sequence content.

\para{Prefix construction.}
We divide the bacteriophage and plasmid training-data subsets into source and validation partitions with balanced numbers of sequence records. Let $\ell$ denote the prefix length in base pairs and $N$ the number of prefixes per run. For the initial 300-bp configuration, we draw one random window of length $\ell=300$ from each eligible source record. We exclude windows containing ambiguous bases or having Shannon entropy below $1.60$ bits per nucleotide, a DUST-masked fraction above $0.05$, or tandem-repeat coverage above $0.20$. Reservoir sampling then selects up to $N=300$ prefixes, with equal target allocation to bacteriophage and plasmid sequences. The prefix-length experiments additionally evaluate 150-bp and 500-bp prefixes, as described in Section~\ref{sec:input-size}.

\para{Repeated generation.}
For each prefix, we query the model with different sampling seeds to obtain up to $K=30$ continuations of length $L=5{,}000$ bp. We analyze only the generated continuation, excluding the input prefix. We retain only complete 5,000-bp continuations consisting exclusively of $\{A,C,G,T\}$; incomplete or invalid generations are resampled within a fixed retry budget. Each standard run therefore targets $NK=9{,}000$ continuations.

\para{Candidate selection and validation.}
The adversary compares continuations generated from the same prefix and identifies clusters of highly similar sequences. The evaluator then aligns every member of an accepted cluster independently against the bacteriophage and plasmid validation subsets. We evaluate extraction using alignment length, nucleotide identity, and the number of matched training records, and apply sequence-complexity and uniqueness filters for final extraction filtering.

\subsection{Cluster-based Extraction}
Algorithm~\ref{alg:extraction} summarizes the attack and its evaluation. For each prefix, we construct a similarity graph over the generated continuations, select clusters satisfying pairwise identity and coverage requirements, and match their members against the validation subset of the training data.

\para{Pairwise convergence.}
We align the continuations of each prefix with minimap2~\cite{bioinformatics2018,minimap2manual}. For a reported alignment $a$, nucleotide identity is computed from matching bases $M_a$, substituted bases $X_a$, inserted bases $I_a$, and deleted bases $D_a$ as
\begin{equation}
\mathrm{ANI}(a)=\frac{M_a}{M_a+X_a+I_a+D_a}.
\label{eq:alignment-identity}
\end{equation}
Insertions and deletions count bases, not gap-opening events. Identity and coverage used for an accepted pair refer to the same alignment.
For a reported alignment $a$ between sequences $u$ and $v$,
let $[b_w(a),e_w(a))$ denote its interval on sequence
$w\in\{u,v\}$. The alignment span on that sequence is
$\Delta_w(a)=e_w(a)-b_w(a)$, measured in base pairs.
This span measures the extent of the aligned region in sequence
coordinates; it may contain mismatches and gaps and therefore
does not represent an uninterrupted exact match.
Insertions and deletions can also cause the spans on the two
sequence sides to differ.

Let $\mathcal{A}(u,v)$ be the set of reported alignments.
For each alignment, we compute its bilateral coverage:
\begin{equation}
c_a(u,v)=\min\left(\frac{\Delta_u(a)}{|u|},
                            \frac{\Delta_v(a)}{|v|}\right).
\end{equation}
We then define maximum continuous alignment coverage (MCAC) as
\begin{equation}
\mathrm{MCAC}(u,v)=
\max_{a\in\mathcal{A}(u,v)}c_a(u,v).
\label{eq:pair-mcac}
\end{equation}
Here, ``continuous'' refers to the coordinate span of an individual
reported alignment rather than the union of disjoint alignments.
Thus, MCAC is the largest fraction that a single reported alignment
covers simultaneously on both sequences. We set it to zero when no
alignment is reported. Pairwise ANI, denoted $\mathrm{ANI}(u,v)$,
is taken from the alignment attaining maximum MCAC, with ties
resolved by ANI, alignment-column count, and mapping quality,
in that order.

MCAC measures the extent of alignment coverage, whereas ANI
measures nucleotide agreement within the selected alignment.
Each continuation forms a vertex in the similarity graph. We connect $u$ and $v$ only when
$\mathrm{MCAC}(u,v)\geq0.70$ and
$\mathrm{ANI}(u,v)\geq0.99$.
For two 5,000-bp continuations, the coverage criterion requires
a single reported alignment spanning at least 3,500 bp on both sequence sides.

\para{Cluster acceptance.}
We identify connected components in the similarity graph and evaluate all sequence pairs within each component. A component is accepted as a candidate cluster if it contains at least $c$ members and every pair satisfies both the ANI and MCAC thresholds. Components failing the pairwise requirements are discarded. We use $c=8$ in the default configuration and evaluate thresholds from 3 to 30 in the cluster-size experiment.

\begin{algorithm}[t]
\caption{Cluster-based training-data extraction and evaluation}
\label{alg:extraction}
\small
\begin{algorithmic}[1]
\Require Model generation interface $f_\theta$, source collection $D_{\mathrm{src}}$
\Require Validation subset $D_{\mathrm{val}}$ (evaluator only)
\Require Prefix count $N$, prefix length $\ell$, continuations per prefix $K$, continuation length $L$, minimum cluster size $c$
\Ensure Candidate clusters $\mathcal C$ and extraction results $\mathcal E$
\State $P\gets\mathrm{FilteredPrefixes}(D_{\mathrm{src}},N,\ell)$
\State $\mathcal C\gets\varnothing$; $\mathcal E\gets\varnothing$
\For{each prefix $x\in P$}
  \State $Y\gets\mathrm{SampleContinuations}(f_\theta,x,K,L)$
  \State Remove prompts and invalid outputs from $Y$
  \State Align every pair in $Y$ and construct graph $G$
  \State Add edges with ANI $\geq0.99$, MCAC $\geq0.70$
  \For{each connected component $C$ of $G$}
    \If{$|C|\geq c$ and every pair passes both tests}
      \State $\mathcal C\gets\mathcal C\cup\{C\}$
    \EndIf
  \EndFor
\EndFor
\For{each $C\in\mathcal C$ and each validation subset $d$}
  \State Align all $y\in C$ to $d\subseteq D_{\mathrm{val}}$
  \State Evaluate match length, identity, and multiplicity
  \State Apply complexity and uniqueness filters
  \State Record extraction results in $\mathcal E$
\EndFor
\State \Return $\mathcal C,\mathcal E$
\end{algorithmic}
\end{algorithm}

\para{Training-data validation.}
For each continuation $y$, the evaluator examines its alignments to each training sequence $t\in D_{\mathrm{val}}$. Length and identity must be supported by the same alignment. Let $\mathcal{A}_{0.99}(y,t)=\{a\in\mathcal{A}(y,t):\mathrm{ANI}(a)\geq0.99\}$, using the per-alignment identity in Equation~\ref{eq:alignment-identity}. We define the qualifying matched length as
\begin{equation}
A(y,t)=\max_{a\in\mathcal A_{0.99}(y,t)}
                \min\{\Delta_y(a),\Delta_t(a)\},
\label{eq:db-mcac}
\end{equation}
measured in base pairs, with $A(y,t)=0$ if no alignment meets the identity threshold. A strong match requires $A(y,t)\geq500$ bp. Thus, a qualifying span is supported on both the continuation and the training record by one alignment whose identity is at least $99\%$. Neither an identity averaged across separate alignments nor a union of disjoint intervals establishes a qualifying long match. The label MCAC 500 or MCAC 1000 below refers to this absolute span in base pairs; pairwise MCAC in Equation~\ref{eq:pair-mcac} is a coverage fraction.

For final extraction filtering, we reject continuations with a low-complexity fraction of at least $0.30$, tandem-repeat fraction of at least $0.30$, a homopolymer of at least 20 bp, five or more high-similarity matched training records, or fewer than 500 uniquely aligned bases. Low-complexity fraction is the proportion of sampled 50-bp windows, at a 10-bp stride, with entropy below $1.2$ bits. Tandem-repeat fraction is the proportion of sequence positions covered by tandem repeats. A uniquely aligned base is covered by a reported high-similarity match to exactly one genomic location in the searched validation subset, with locations distinguished by sequence identifiers and coordinate intervals. These complexity filters operate on the full 5,000-bp continuation.  Section~\ref{sec:source-overlap} additionally checks representative surviving fragments. 

\subsection{Evaluation Metrics}
We distinguish candidate-cluster matching criteria from the stricter final uniquely matched extraction criterion, which additionally requires a single reported matched record and passage of the final extraction filters. We report five criteria, with two length thresholds for the final criterion. Results are grouped by experiment, cluster, and validation subset, so multiple continuations from the same cluster matching the same subset contribute one result.

For a continuation $y$ and validation subset $d$, let $A_{y,d}=\max_{t\in d}A(y,t)$, and let $H_{y,d}$ count the distinct records in $d$ with a reported high-similarity match to $y$. The latter counts records, not alignments or copies in the full training corpus. Let $F_{y,d}=1$ indicate passage of all final extraction filters. For cluster $C$, define $A_{C,d}=\max_{y\in C}A_{y,d}$ and $H_{C,d}=\min_{y\in C:\,A_{y,d}\geq500}H_{y,d}$. If the restricted set is empty, set $H_{C,d}=+\infty$.

\begin{enumerate}
\item \textbf{MCAC 500:} $A_{C,d}\geq500$ bp.
\item \textbf{MCAC 1000:} $A_{C,d}\geq1{,}000$ bp.
\item \textbf{MCAC 1000 + unique 10:} $A_{C,d}\geq1{,}000$ and $H_{C,d}\leq10$.
\item \textbf{MCAC 1000 + unique 5:} $A_{C,d}\geq1{,}000$ and $H_{C,d}\leq5$.
\item \textbf{Unique matched ($\geq\tau$ bp):} there exists $y\in C$ such that $A_{y,d}\geq\tau$, $H_{y,d}=1$, and $F_{y,d}=1$. We report this criterion at $\tau=500$ and $\tau=1{,}000$ bp.
\end{enumerate}
Criteria 1–-4 are cluster-level screening criteria, for which length and multiplicity statistics may come from different continuations. The final Unique matched criterion instead requires one continuation to satisfy the length, unique-match, and filtering requirements jointly. 

For each criterion, let $S$ be the number of candidate clusters satisfying it in at least one validation subset and $Q$ the total number of candidate clusters meeting the size threshold. We define the cluster-level attack success rate as
\begin{equation}
\mathrm{ASR}_{\mathrm{cluster}}=S/Q.
\label{eq:asr}
\end{equation}
A cluster with successful matches in both the bacteriophage and plasmid subsets is counted once in the combined results. Repeated extraction of the same training-sequence region in different experiments remains in $S$. To quantify the diversity of extracted sequences, we also report the number of distinct matched training records and the number of extracted regions after merging overlapping intervals within each record.

\section{Evaluation}
\label{sec:evaluation}
We evaluate training-sequence extraction from Evo 2, examining extraction success, prefix length, and the relationship between cluster size and extraction yield. All attack success rates in this section are cluster-level rates, $\mathrm{ASR}_{\mathrm{cluster}}$, as defined in Section~\ref{sec:method}.

\subsection{Overall Results}
\para{Experimental setup.}
We target the Evo 2 7B model~\cite{nature2026-1}. Each standard run uses 300 prefixes and targets 30 continuations of 5,000 bases per prefix. We use a sampling temperature of $0.35$ and top-$p$ of $0.92$, with generation parallelized across 16 GPUs.

The main experiments comprise 152 runs with 300-bp prefixes and 152 with 500-bp prefixes, with 242 runs executed on NVIDIA A100 GPUs and 62 on NVIDIA H100 GPUs. An early-stopping mechanism is enabled in 105 runs: generation is periodically checked for short repetitive motifs, and an attempt is terminated when such a pattern is detected. The remaining 199 runs proceed without this mechanism. We aggregate all 304 runs when reporting overall extraction results, targeting $304\times300\times30=2{,}736{,}000$ continuations before retries. This broad repeated-generation design supports stable comparison across prefix lengths, cluster sizes, and generation procedures.

\para{Default-setting extraction.}
Table~\ref{tab:overall} pools both prefix lengths at the cluster-size threshold $c=8$, retaining repeated extraction events across experiments. Across the 304 runs, the default setting yields 888 eligible candidate clusters and 85 MCAC-500 extraction events, giving a cluster-level ASR of $9.57\%$. Raising the aligned-span requirement to 1,000 bp yields 45 events ($5.07\%$). The MCAC 1000 + unique 10 and MCAC 1000 + unique 5 criteria yield 32 and 26 events, respectively, with ASRs of $3.60\%$ and $2.93\%$. After final filtering, the Unique matched criterion yields 22 events with aligned spans of at least 500 bp ($2.48\%$), including 15 of at least 1,000 bp ($1.69\%$). The remaining seven events have spans in $[500,1{,}000)$ bp, corresponding to an ASR of $0.79\%$ and accounting for $31.8\%$ of the final uniquely matched extractions. All seven are plasmid extractions obtained with 500-bp prefixes.

\begin{table}[t]
\centering
\caption{Extraction results at the default cluster-size threshold ($c=8$). Entries are counts (ASR in percent).}
\label{tab:overall}
\small
\setlength{\tabcolsep}{3pt}
\begin{tabular}{@{}lrrr@{}}
\toprule
Criterion & 300 bp & 500 bp & Combined \\
\midrule
Runs & 152 & 152 & 304 \\
Eligible clusters & 418 & 470 & 888 \\
\midrule
MCAC 500 & 25 (5.98) & 60 (12.77) & 85 (9.57) \\
MCAC 1000 & 22 (5.26) & 23 (4.89) & 45 (5.07) \\
\quad + unique 10 & 17 (4.07) & 15 (3.19) & 32 (3.60) \\
\quad + unique 5 & 14 (3.35) & 12 (2.55) & 26 (2.93) \\
Unique matched ($\geq500$ bp) & 9 (2.15) & 13 (2.77) & 22 (2.48) \\
Unique matched ($\geq1{,}000$ bp) & 9 (2.15) & 6 (1.28) & 15 (1.69) \\
\bottomrule
\end{tabular}
\par\smallskip
\begin{minipage}{\columnwidth}\footnotesize
ASR denotes $\mathrm{ASR}_{\mathrm{cluster}}$, the number of successful clusters divided by the number of eligible clusters. Repeated extractions of the same training-sequence region are counted separately. The unique 10 and unique 5 criteria require MCAC $\geq1{,}000$ bp.
Both Unique matched rows require the same continuation to meet the indicated length threshold, match a single record in the searched validation subset, and pass all final filters. Their length thresholds are nested.
\end{minipage}
\end{table}

Across the 304 runs, 71 contain at least one MCAC-500 event, showing that recovery is distributed across the run cohort rather than concentrated in a single experiment.

\para{Repeated recovery and distinct extracted regions.}
We distinguish repeated extraction events from distinct extracted training regions. The 85 MCAC-500 events comprise 69 plasmid and 16 bacteriophage hits, matching 70 distinct training records and covering 70 extracted sequence regions. To count distinct regions, we group hits by training subset and accession, then merge connected overlapping target intervals irrespective of alignment strand. For example, five recoveries map to overlapping intervals of phage accession \texttt{NC\_049950.1} and contribute one region to the total. The longest MCAC is 5,000 bp, spanning the full continuation length. Event counts measure extraction yield, while merged-region counts provide the corresponding locus-level view.

The same counting rule applies when the overlap is short relative to the recovered intervals. In one case, intervals of 4,827 bp and 5,000 bp overlap by 127 bp and are therefore counted as one extracted region.

Applying this interval-merging rule to the final unique extractions gives 14 distinct regions in 14 training records at the 500-bp threshold, including seven regions in seven records at the 1,000-bp threshold. The seven additional regions each have an extracted span in $[500,1{,}000)$ bp. Thus, the shorter unique extractions recover additional training regions beyond those represented by the longer matches.

As a supplementary fragment-level diagnostic, we evaluate 284 retained query-side fragments supporting the default-setting matches at the 500-bp threshold. We apply the same entropy-window, tandem-repeat, and homopolymer functions directly to each fragment. Each of the 22 clusters has at least one fragment passing all three complexity thresholds, confirming that the retained matches include locally complex recovered content.
\subsection{Effect of Prefix Length}
\label{sec:input-size}
We construct 500-bp prefixes by extending each 300-bp prefix with the next 200 nucleotides from the same source sequence, preserving its start coordinate. For the 150-bp experiments, we select 300-bp prefixes whose continuations form a cluster of at least four members, then use the front and back halves of each prefix as separate prompts.

Across the evaluated runs, the 300-bp and 500-bp configurations yield 25 and 60 MCAC-500 events, with cluster-level ASRs of $5.98\%$ and $12.77\%$, respectively. The difference is concentrated in shorter matches: MCAC-1000 events remain similar (22 versus 23). Final uniquely matched extractions of at least 500 bp number nine ($2.15\%$) and 13 ($2.77\%$), whereas those of at least 1,000 bp number nine and six, respectively. Thus, all nine uniquely matched extractions from 300-bp prefixes reach 1,000 bp, while seven of the 13 from 500-bp prefixes fall in $[500,1{,}000)$ bp. Across all MCAC-500 events, 37 of 60 in the 500-bp group fall below 1,000 bp, compared with three of 25 in the 300-bp group.

To compare all three prefix lengths, we use the 17 run-level seeds with 150-bp experiments and their corresponding 300- and 500-bp runs. Table~\ref{tab:input-size} applies the same $c=8$, MCAC-500 criterion to each group. These seeds identify experimental runs and are distinct from the sampling seeds used for individual continuations. The 150-bp runs produce one positive cluster out of 32 ($3.13\%$), compared with eight out of 52 ($15.38\%$) for 300 bp and 12 out of 51 ($23.53\%$) for 500 bp. Within these 17 matched run-level seeds, both the number of positive clusters and the cluster-level ASR increase with prefix length. The 150-bp cohort is constructed by splitting selected 300-bp prefixes and uses two prompts per selected prefix.

\begin{table}[t]
\centering
\caption{Prefix-length comparison on 17 shared run-level seeds. All groups use $c=8$ and the MCAC-500 criterion; ASR denotes the cluster-level rate.}
\label{tab:input-size}
\small
\setlength{\tabcolsep}{5pt}
\begin{tabular}{@{}lrrrr@{}}
\toprule
Prefix & Clusters & Hits & ASR (\%) & Hit runs \\
\midrule
150 bp$^*$ & 32 & 1 & 3.13 & 1/17 \\
300 bp & 52 & 8 & 15.38 & 5/17 \\
500 bp & 51 & 12 & 23.53 & 9/17 \\
\bottomrule
\end{tabular}
\par\smallskip
\begin{minipage}{\columnwidth}\footnotesize
$^*$The 150-bp prefixes are the front and back halves of 300-bp prefixes whose continuations form a cluster of at least four members.
\end{minipage}
\end{table}

\begin{table*}[!b]
\centering
\caption{Extraction results at different cluster-size thresholds. Each entry is a positive cluster count (ASR in percent). Clusters matching both training subsets are counted once.}
\label{tab:cluster-size}
\small
\setlength{\tabcolsep}{7pt}
\begin{tabular}{@{}rrrrrrrr@{}}
\toprule
Threshold $c$ & Eligible clusters & MCAC 500 & MCAC 1000 & + unique 10 & + unique 5 & \multicolumn{2}{c}{Unique matched} \\
 & & & & & & $\geq500$ bp & $\geq1{,}000$ bp \\
\midrule
3 & 14,535 & 380 (2.61) & 107 (0.74) & 67 (0.46) & 59 (0.41) & 71 (0.49) & 31 (0.21) \\
5 & 3,151 & 153 (4.86) & 60 (1.90) & 43 (1.36) & 37 (1.17) & 36 (1.14) & 20 (0.63) \\
8 & 888 & 85 (9.57) & 45 (5.07) & 32 (3.60) & 26 (2.93) & 22 (2.48) & 15 (1.69) \\
10 & 607 & 64 (10.54) & 40 (6.59) & 29 (4.78) & 24 (3.95) & 20 (3.29) & 13 (2.14) \\
15 & 334 & 42 (12.57) & 32 (9.58) & 25 (7.49) & 20 (5.99) & 15 (4.49) & 12 (3.59) \\
20 & 216 & 32 (14.81) & 27 (12.50) & 21 (9.72) & 16 (7.41) & 9 (4.17) & 8 (3.70) \\
25 & 144 & 23 (15.97) & 19 (13.19) & 15 (10.42) & 13 (9.03) & 7 (4.86) & 7 (4.86) \\
30 & 86 & 14 (16.28) & 10 (11.63) & 7 (8.14) & 7 (8.14) & 3 (3.49) & 3 (3.49) \\
\bottomrule
\end{tabular}
\par\smallskip
\begin{minipage}{0.96\textwidth}\footnotesize
The unique 10 and unique 5 columns are composite cluster-level screens with MCAC $\geq1{,}000$ bp. The final two columns require one continuation to satisfy the indicated length threshold, match a single record, and pass all final filters. Increasing $c$ selects progressively smaller, nested subsets of the generated clusters.
\end{minipage}
\end{table*}

\subsection{Effect of Cluster Size}
\label{sec:cluster-size}
We next vary the minimum cluster size while keeping the generated continuations and pairwise alignment criteria fixed. This experiment covers 304 runs and 14,535 clusters of at least three members.

\begin{figure*}[t]
\centering
\includegraphics[width=\textwidth]{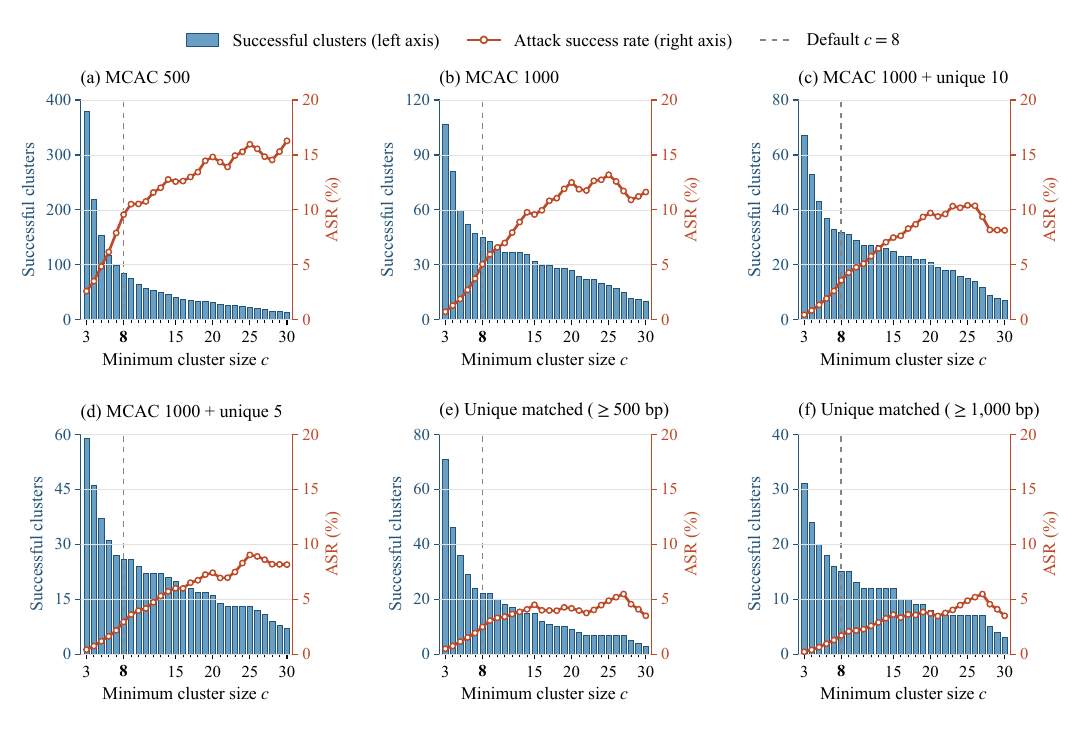}
\caption{Extraction results across all integer minimum cluster-size thresholds $c\in[3,30]$. Blue bars show successful cluster counts (left axes), and orange lines with circular markers show ASR (right axes) under each extraction criterion. At each threshold, ASR is the successful cluster count divided by the number of clusters with at least $c$ members. Dashed vertical lines mark the default $c=8$. All panels use the same ASR scale; count scales vary by panel.}
\label{fig:cluster-size}
\end{figure*}

Figure~\ref{fig:cluster-size} shows successful cluster counts and cluster-level ASR at every integer threshold from 3 to 30, while Table~\ref{tab:cluster-size} reports representative thresholds together with the eligible cluster counts. Increasing $c$ reduces the number of candidates sharply: 14,535 clusters are eligible at $c=3$, 888 at $c=8$, and 86 at $c=30$. The MCAC-500 positive count falls from 380 to 85 to 14, respectively, while ASR rises from $2.61\%$ to $9.57\%$ to $16.28\%$. Stronger within-prefix convergence therefore increases the concentration of successful extractions while selecting a smaller candidate set. Relative to $c=3$, the default $c=8$ retains $6.11\%$ of candidates and $22.37\%$ of MCAC-500 positives.

The stricter criteria show a nonmonotonic relationship at the highest thresholds. MCAC-1000 ASR rises to $13.19\%$ at $c=25$ and decreases to $11.63\%$ at $c=30$. ASR under the Unique matched criterion at both length thresholds similarly decreases from $4.86\%$ to $3.49\%$. For these criteria, $c=25$ yields a higher ASR than $c=30$ while retaining more positive clusters. The difference between the two uniquely matched extraction counts narrows from 40 events at $c=3$ to seven at $c=8$, and disappears at $c\geq25$: all uniquely matched extractions retained at these highest thresholds reach 1,000 bp.



\section{Discussion}
\label{sec:discussion}
We further examine the relationship between perplexity and extraction, the effect of suppressing repetitive generation, extraction from messenger RNA (mRNA) sequences, and overlap between recovered sequences and the attacker's source collection.

\subsection{Perplexity as an Extraction Signal}
Perplexity has been used to prioritize extraction candidates in language models~\cite{usenix2021}. We examine this signal through an auxiliary experiment using greedy decoding, which selects the most probable next nucleotide at each generation step. The initial analysis covers 27 runs, each using 300 prefixes of length 300 bp and generating one 5,000-bp continuation per prefix. This auxiliary analysis supplies a likelihood-based ranking signal alongside the output-only attack. For each complete valid continuation, we compute perplexity as the exponential of its mean next-nucleotide negative log-likelihood, using preceding generated nucleotides as context. The scored region therefore contains 5,000 nucleotides; the original prefix is not prepended for scoring. Lower perplexity indicates higher average probability assigned to the evaluated nucleotides.

The 27 runs target $27\times300=8{,}100$ prefix-level outputs. Generation logs identify 8,049 valid outputs and 51 terminal failures after the allowed retries. Among valid continuations, 147 satisfy MCAC 500 and 86 satisfy MCAC 1000, corresponding to $1.83\%$ and $1.07\%$ of the continuations, respectively. The median perplexity is $1.079$ for continuations satisfying MCAC 500 and $1.008$ for the remaining continuations, so successful matches occupy the higher-perplexity side of this cohort. Selecting the lowest-perplexity continuation within each run yields 27 candidates, none satisfying MCAC 500, and 26 violating at least one sequence-complexity or repeat threshold. The result identifies a clear selection effect: the most predictable continuations are dominated by repetitive sequences, whereas successful extraction candidates are better associated with higher perplexity and explicit complexity filtering.

\para{Threshold-based classification.}
We further evaluate perplexity as a binary classifier using the complete labelled collection of 11,100 scored terminal outcomes from 37 runs in total, including the 27 runs described above and 10 additional runs from other experiments, with 300 prefix-level outcomes per run. The collection contains 39 supplied positive labels and 11,061 supplied negative labels. These records are not all retry-level model calls. Associating each score with its terminal generation outcome accounts for the 51 failed outputs above: they remain in this collection with the last recorded score and a supplied negative label. The additional ten runs contribute 3,000 outcomes, comprising 2,939 valid outputs and 61 failures. Altogether, the collection contains 10,988 valid outputs and 112 failures, all of the latter labelled negative. Thus, its metrics describe selection over prefix-level terminal outcomes, including generation failures, rather than over valid continuations alone. This collection contains 39 final extraction events with MCAC of at least 500 bp, including 28 uniquely matched extractions of at least 1,000 bp. These counts concern individual scored terminal outcomes and we use this collection to assess whether perplexity can identify successful continuations. 

Let $p$ denote a continuation's perplexity and $t$ a decision threshold. We evaluate both $p\leq t$ and $p\geq t$, allowing either lower or higher perplexity to indicate extraction. For each extraction-length criterion, we exhaustively search all distinct score cutoffs and report precision, recall, $F_1$, and AUROC over the 11,100 terminal outcomes. This directly quantifies the ranking value of perplexity for the two extraction criteria.

\begin{table}[t]
\centering
\caption{Perplexity-threshold classification over 11,100 prefix-level terminal outcomes, including 112 generation failures labelled negative. For each direction and length criterion, the displayed cutoff maximizes both precision and $F_1$ on this labelled collection. Precision, recall, and $F_1$ are percentages.}
\label{tab:ppl-classification}
\small
\setlength{\tabcolsep}{3pt}
\begin{tabular}{@{}lrrrrr@{}}
\toprule
Rule & TP & FP & Precision & Recall & $F_1$ \\
\midrule
\multicolumn{6}{@{}l}{Final extraction $\geq500$ bp: 39 positives} \\
$p\leq1.2979$ & 39 & 10,784 & 0.36 & 100.00 & 0.72 \\
$p\geq1.1903$ & 16 & 825 & 1.90 & 41.03 & 3.64 \\
\midrule
\multicolumn{6}{@{}l}{Uniquely matched extraction $\geq1{,}000$ bp: 28 positives} \\
$p\leq1.2465$ & 28 & 10,502 & 0.27 & 100.00 & 0.53 \\
$p\geq1.1903$ & 10 & 831 & 1.19 & 35.71 & 2.30 \\
\bottomrule
\end{tabular}
\end{table}

Table~\ref{tab:ppl-classification} shows that the higher-perplexity rule achieves higher precision and $F_1$. At $p\geq1.1903$, 841 terminal outcomes are selected, of which 16 are final extractions: precision is $1.90\%$, recall is $41.03\%$, and $F_1$ is $3.64\%$. Requiring at least 1,000 bp gives ten true positives among the same candidates, with precision of $1.19\%$ and $F_1$ of $2.30\%$. The higher-perplexity ranking has an area under the receiver operating characteristic curve (AUROC) of $0.833$ for final extraction and $0.800$ for uniquely matched extraction of at least 1,000 bp.  

The higher-perplexity rule raises precision from the supplied positive prevalence of approximately $0.35\%$ to $1.90\%$, while retaining 41.03\% of the final-extraction events and 35.71\% of the uniquely matched events at the 1,000-bp threshold. Perplexity therefore provides a measurable ranking signal, whereas convergence-based clustering remains the direct output-only selection mechanism because it uses only observable agreement among repeated continuations. Independently of the ranking direction, the 39 final extractions, including 28 reaching 1,000 bp, show that greedy generation can produce continuations containing substantial training-sequence content.

\subsection{Effect of Early Stopping}
Early stopping detects short-period repetition during generation. After each 100-bp extension, it examines up to the most recent 500 bp and terminates an attempt if that window consists entirely of a repeating motif of length 1--9 bp. The model then resamples the continuation, and only complete, valid 5,000-bp outputs enter the subsequent analysis. Direct generation requests the full continuation without this intermediate repetition check. Both groups use the same clustering and extraction criteria.

Table~\ref{tab:early-stop} compares 105 runs with early stopping and 199 with direct generation at $c=8$. Their overall MCAC-500 cluster-level ASRs are $9.47\%$ and $9.64\%$, respectively. Under the Unique matched criterion, the corresponding ASRs are $2.37\%$ and $2.55\%$ at the 500-bp threshold, and $1.78\%$ and $1.64\%$ at the 1,000-bp threshold. The direction of the difference also varies with prefix length: for 300-bp prefixes, MCAC-500 ASR is $6.43\%$ with early stopping and $5.67\%$ with direct generation; for 500-bp prefixes, it is $12.57\%$ and $12.87\%$, respectively. We observe no consistent improvement in extraction success for the early-stopping runs compared with the direct-generation runs. 

Under our implementation, generating a complete 5,000-bp continuation on an NVIDIA A100 GPU takes approximately 100--110 s without early stopping and 140--150 s with early stopping. The latter is about $1.4\times$ the runtime, or approximately $40\%$ longer. These timings indicate additional generation cost without a consistent increase in cluster-level extraction success in the evaluated runs.

\begin{table}[t]
\centering
\caption{Extraction with early stopping and direct generation at $c=8$. Extraction entries are counts (cluster-level ASR in percent), pooled over 300-bp and 500-bp prefixes.}
\label{tab:early-stop}
\small
\setlength{\tabcolsep}{2pt}
\begin{tabular}{@{}lrr@{}}
\toprule
Criterion & Early stopping & Direct generation \\
\midrule
Runs & 105 & 199 \\
Eligible clusters & 338 & 550 \\
\midrule
MCAC 500 & 32 (9.47) & 53 (9.64) \\
MCAC 1000 & 16 (4.73) & 29 (5.27) \\
\quad + unique 10 & 10 (2.96) & 22 (4.00) \\
\quad + unique 5 & 9 (2.66) & 17 (3.09) \\
Unique matched ($\geq500$ bp) & 8 (2.37) & 14 (2.55) \\
Unique matched ($\geq1{,}000$ bp) & 6 (1.78) & 9 (1.64) \\
\bottomrule
\end{tabular}
\end{table}

\subsection{Extraction from mRNA Sequences}
We used the mRNA cohort to extend the attack beyond phage and plasmid sequences. The mRNA collection is associated with the released OpenGenome2 mRNA component~\cite{arc2025opengenome2}. The experiment extends the extraction evaluation to disjoint source and validation partitions of eukaryotic transcript records, broadening the sequence domains tested beyond bacteriophage and plasmid data. We also apply the extraction procedure to mRNA training sequences, using ten runs with 300-bp prefixes and ten with 500-bp prefixes. Each run targets 300 prefixes and 30 continuations of 5,000 bp per prefix. With the same pairwise alignment requirements and minimum cluster size $c=8$, the two prefix-length groups produce 64 and 51 eligible clusters, respectively. The 300-bp group yields three MCAC-500 extraction events ($4.69\%$), including two satisfying MCAC 1000 ($3.13\%$). The 500-bp group yields one MCAC-500 event ($1.96\%$). Across both groups, four of 115 clusters satisfy MCAC 500, giving an overall cluster-level ASR of $3.48\%$; two satisfy both MCAC 1000 + unique 10 and MCAC 1000 + unique 5 ($1.74\%$).

One mRNA cluster satisfies the Unique matched criterion with a maximum aligned span of 531 bp. Two additional clusters contain matches exceeding 1,000 bp but are removed by the stated multiplicity or complexity filters. These results extend the observed extraction behavior beyond bacteriophage and plasmid sequences.

\subsection{Controlling for Homology in the Source Data}
\label{sec:source-overlap}
At the default $c=8$, the source-homology control retains 11 of the 22 Unique matched events at a $99\%$ source-exclusion threshold (eight phage and three plasmid), and four at $98\%$ (three phage and one plasmid). The four surviving events map to two reference regions after merging overlapping intervals within each record. Thus, 11 of the 22 final extraction events persist after removing near-identical homologues from the corresponding source collection.

To quantify the contribution of source homology, we re-align strong regions matching the phage validation subset against the phage source partition, and plasmid regions against the plasmid source partition. Each region entering this control has validation identity $\geq99\%$ and an individual alignment spanning at least 500 bp on both sequence sides. The resulting counts measure the extraction signal that remains after within-category source screening.

A source target is counted as a homology explanation when its reported alignments jointly meet an identity threshold $\eta$ and cover at least $\min(500,0.9L)$ bp on both the query and target sides, where $L$ is the extracted-region length. Covered lengths use the respective unions of aligned intervals for that source target, and identity and coverage refer to the same target. The rule accepts distributed source alignments and allows near-complete coverage for regions close to 500 bp. A cluster survives when at least one individually validated region has no source target meeting the rule; multiple continuations or regions in a cluster still contribute one event.

\begin{table*}[t]
\centering
\caption{Representative alignments for the four default-setting events surviving the $98\%$ within-category source control. Each row uses one continuation and one validation alignment. Target intervals are zero-based and half-open.}
\label{tab:source-survivors}
\small
\setlength{\tabcolsep}{5pt}
\begin{tabular}{@{}lllrrl@{}}
\toprule
Event & Validation accession & Target interval & Span (bp) & Identity (\%) & Source search \\
\midrule
Event 1$^*$ & NC\_031129.1 & $[58582,62222)$ & 3,640 & 99.5604 & No reported hit \\
Event 2$^*$ & NC\_031129.1 & $[58691,62222)$ & 3,531 & 99.5185 & No reported hit \\
Event 3 & NC\_031129.1 & $[58477,61983)$ & 3,506 & 99.8859 & No reported hit \\
Event 4$^*$ & NZ\_CP040809.1 & $[13504,14021)$ & 517 & 99.2263 & No reported hit \\
\bottomrule
\end{tabular}
\par\smallskip
\begin{minipage}{0.96\textwidth}\footnotesize
$^*$500-bp-prefix run; the remaining run uses a 300-bp prefix. Within each cluster, the representative has the greatest surviving bilateral span, with ties resolved by identity. The first three rows overlap within one phage record; the fourth lies in a plasmid record. 
\end{minipage}
\end{table*}

Table~\ref{tab:source-survivors} provides individual alignment evidence for these four events. As a supplementary check against dilution of local repeats by the full continuation, we apply the same complexity functions to the query-side fragments supporting the displayed alignments. All four have a low-complexity fraction of zero; tandem-repeat fractions range from $1.74\%$ to $2.25\%$, and their longest homopolymers span five or six bases. Each representative therefore passes the three stated complexity thresholds on the recovered fragment itself.

Across all cluster sizes ($c\geq3$), we apply the control to clusters with final-retained strong content, irrespective of validation-target uniqueness. The reported cohort contains 84 distinct evaluable clusters: all 71 satisfying the Unique matched ($\geq500$ bp) criterion in Table~\ref{tab:cluster-size}, together with 13 additional clusters whose retained strong sequences have two to four high-similarity validation targets. This difference from Table~\ref{tab:cluster-size} reflects the inclusion of multiple-record matches and the requirement for an individually qualifying strong region.

At $\eta=99\%$, 21 of 27 phage-associated clusters and five of 57 plasmid-associated clusters remain after source-homology screening.

Lowering the source-exclusion threshold to $\eta=98\%$ applies a stricter control by also removing regions with lower-identity source homologues. Under this criterion, 15 clusters remain: 13 phage and two plasmid clusters. The same 15 survive thresholds of $97\%$, $96\%$, and $95\%$; their surviving strong content has no reported source hit, rather than an alignment narrowly below the identity cutoff. Eleven of the 13 phage clusters retain an individually validated alignment span of at least 1,000 bp.

For direct comparison with Table~\ref{tab:cluster-size}, restricting the control to the 71 evaluable uniquely matched clusters yields 24 surviving clusters at $\eta=99\%$ (19 phage and five plasmid). All 15 clusters surviving the $98\%$--$95\%$ thresholds belong to this uniquely matched subset. The default-setting results reported at the start of this section restrict this cohort to $c=8$.

Source-side screening leaves a strong phage-enriched survivor set. At $98\%$--$95\%$, 15 clusters remain, including 13 phage clusters, and 11 of those retain an individually validated span of at least 1,000 bp. The result shows that long, high-identity extraction candidates persist after a stringent within-category homology control.

\subsection{Security Implications}
The experiments demonstrate that a generation interface can expose substantial sequence content matching a genomic model's training data, and that repeated-output convergence can identify candidates without likelihood access. The same attack therefore provides a practical audit for genomic models trained on restricted sequence collections. Our benchmark uses publicly released sequences, while the attack mechanism applies directly to any training domain exposed through the generation interface.


\FloatBarrier
\section{Conclusion}
\label{sec:conclusion}

We presented an output-only black-box attack that extracts training-sequence content from Evo 2 through repeated generation from short nucleotide prefixes. The method identifies candidate continuations using convergence measured through pairwise nucleotide identity and alignment coverage, then validates recovered content with aligned length, multiplicity, and complexity criteria. Evaluation against bacteriophage and plasmid training data demonstrates recovery of segments spanning up to 5,000 bp. At the default cluster-size threshold, 85 of 888 candidate clusters contain continuations matching at least 500 bp, yielding a cluster-level attack success rate of $9.57\%$. The final Unique matched criterion retains 22 extraction events of at least 500 bp, including 15 of at least 1,000 bp; merging overlapping recoveries gives 14 and seven distinct extracted regions, respectively. Source-side homology screening retains 11 of the 22 default-setting events under the $99\%$ exclusion threshold and four under the stricter $98\%$ threshold, with the latter mapping to two reference regions.

In the evaluated runs, longer prefixes yield more matches primarily below 1,000 bp, while larger clusters concentrate successful extractions among fewer retained candidates. Selecting the lowest-perplexity continuations predominantly favors repetitive sequences in the greedy-generation experiment. Extraction remains observable under both early-stopping and direct-generation settings. Applying the procedure to mRNA training sequences additionally yields a uniquely matched extraction spanning 531 bp. These findings support convergence among generated continuations as an output-only signal for identifying extraction candidates.

Together, the prefix-length, cluster-size, perplexity, early-stopping, mRNA, and source-control experiments show that convergence-based extraction is measurable across generation settings and sequence domains. The attack provides a concrete privacy-audit capability for genomic foundation models.

\bibliographystyle{IEEEtran}
\bibliography{bib}

@IEEEtranBSTCTL{IEEEexample:BSTcontrol,
  CTLuse_forced_etal = "yes",
  CTLmax_names_forced_etal = "15",
  CTLnames_show_etal = "3",
  CTLdash_repeated_names = "no"
}

@inbook{BCB2026,
author = {Nemecek, Alexander and Li, Wenbiao and Jiang, Xiaoqian and Vaidya, Jaideep and Ayday, Erman},
title = {Quantifying Memorization and Privacy Risks in Genomic Language Models},
year = {2026},
isbn = {9798400726538},
publisher = {Association for Computing Machinery},
url = {https://doi.org/10.1145/3807503.3819470},
booktitle = {Proceedings of the 17th ACM International Conference on Bioinformatics, Computational Biology and Health Informatics},
}

@article{bioinformatics2018,
    author = {Li, Heng},
    title = {Minimap2: pairwise alignment for nucleotide sequences},
    journal = {Bioinformatics},
    volume = {34},
    number = {18},
    year = {2018},
    issn = {1367-4803},
    doi = {10.1093/bioinformatics/bty191},
}

@article{bioinformatics2021-1,
    author = {Ji, Yanrong and Zhou, Zhihan and Liu, Han and Davuluri, Ramana V},
    title = {DNABERT: pre-trained Bidirectional Encoder Representations from Transformers model for DNA-language in genome},
    journal = {Bioinformatics},
    volume = {37},
    number = {15},
    year = {2021},
    month = {08},
    issn = {1367-4803},
    doi = {10.1093/bioinformatics/btab083},
}

@article{bioinformatics2021-2,
    author = {Bu, Diyue and Wang, Xiaofeng and Tang, Haixu},
    title = {Haplotype-based membership inference from summary genomic data},
    journal = {Bioinformatics},
    volume = {37},
    number = {Supplement\_1},
    year = {2021},
    issn = {1367-4803},
    doi = {10.1093/bioinformatics/btab305},
}

@inproceedings{COLM2024,
title={Do Membership Inference Attacks Work on Large Language Models?},
author={Michael Duan and Anshuman Suri and Niloofar Mireshghallah and Sewon Min and Weijia Shi and Luke Zettlemoyer and Yulia Tsvetkov and Yejin Choi and David Evans and Hannaneh Hajishirzi},
booktitle={First Conference on Language Modeling},
year={2024},
url={https://openreview.net/forum?id=av0D19pSkU}
}

@inproceedings{ICLR2023,
title={Quantifying Memorization Across Neural Language Models},
author={Nicholas Carlini and Daphne Ippolito and Matthew Jagielski and Katherine Lee and Florian Tramer and Chiyuan Zhang},
booktitle={The Eleventh International Conference on Learning Representations },
year={2023},
url={https://openreview.net/forum?id=TatRHT_1cK}
}

@inproceedings{ICLR2026,
title={GeneBreaker: Jailbreak Attacks against {DNA} Language Models with Pathogenicity Guidance},
author={ZAIXI ZHANG and Zhenghong Zhou and Ruofan Jin and Le Cong and Mengdi Wang},
booktitle={The Fourteenth International Conference on Learning Representations},
year={2026},
url={https://openreview.net/forum?id=C5OIolrNJd}
}

@Article{nature2026-1,
author={Brixi, Garyk and Durrant, Matthew G. and Ku, Jerome and Naghipourfar, Mohsen and Poli, Michael and Sun, Gwanggyu and Brockman, Greg and Chang, Daniel and Fanton, Alison and Gonzalez, Gabriel A. and King, Samuel H. and Li, David B. and Merchant, Aditi T. and Nguyen, Eric and Ricci-Tam, Chiara and Romero, David W. and Schmok, Jonathan C. and Taghibakhshi, Ali and Vorontsov, Anton and Yang, Brandon and Deng, Myra and Gorton, Liv and Nguyen, Nam and Wang, Nicholas K. and Pearce, Michael T. and Simon, Elana and Adams, Etowah and Amador, Zachary J. and Ashley, Euan A. and Baccus, Stephen A. and Dai, Haoyu and Dillmann, Steven and Ermon, Stefano and Guo, Daniel and Herschl, Michael H. and Ilango, Rajesh and Janik, Ken and Lu, Amy X. and Mehta, Reshma and Mofrad, Mohammad R. K. and Ng, Madelena Y. and Pannu, Jaspreet and R{\'e}, Christopher and St. John, John and Sullivan, Jeremy and Tey, Joseph and Viggiano, Ben and Zhu, Kevin and Zynda, Greg and Balsam, Daniel and Collison, Patrick and Costa, Anthony B. and Hernandez-Boussard, Tina and Ho, Eric and Liu, Ming-Yu and McGrath, Thomas and Powell, Kimberly and Pinglay, Sudarshan and Burke, Dave P. and Goodarzi, Hani and Hsu, Patrick D. and Hie, Brian L.},
title={Genome modelling and design across all domains of life with Evo 2},
journal={Nature},
year={2026},
volume={652},
number={8112},
issn={1476-4687},
doi={10.1038/s41586-026-10176-5},
}

@Article{nature2026-2,
author={Avsec, {\v{Z}}iga and Latysheva, Natasha and Cheng, Jun and Novati, Guido and Taylor, Kyle R. and Ward, Tom and Bycroft, Clare and Nicolaisen, Lauren and Arvaniti, Eirini and Pan, Joshua and Thomas, Raina and Dutordoir, Vincent and Perino, Matteo and De, Soham and Karollus, Alexander and Gayoso, Adam and Sargeant, Toby and Mottram, Anne and Wong, Lai Hong and Drot{\'a}r, Pavol and Kosiorek, Adam and Senior, Andrew and Tanburn, Richard and Applebaum, Taylor and Basu, Souradeep and Hassabis, Demis and Kohli, Pushmeet},
title={Advancing regulatory variant effect prediction with AlphaGenome},
journal={Nature},
year={2026},
volume={649},
number={8099},
doi={10.1038/s41586-025-10014-0},
}

@Article{nature2026-3,
author={Knolle, Moritz A. and Menten, Martin J. and Jungmann, Friederike and Meissen, Felix and Glocker, Ben and Rueckert, Daniel and Kaissis, Georgios},
title={Disparate privacy risks from medical AI},
journal={Nature},
year={2026},
month={Aug},
volume={656},
number={8126},
issn={1476-4687},
doi={10.1038/s41586-026-10688-0},
url={https://doi.org/10.1038/s41586-026-10688-0}
}

@Article{naturemethod2025,
author={Dalla-Torre, Hugo and Gonzalez, Liam and Mendoza-Revilla, Javier and Lopez Carranza, Nicolas and Grzywaczewski, Adam Henryk and Oteri, Francesco and Dallago, Christian and Trop, Evan and de Almeida, Bernardo P. and Sirelkhatim, Hassan and Richard, Guillaume and Skwark, Marcin and Beguir, Karim and Lopez, Marie and Pierrot, Thomas},
title={Nucleotide Transformer: building and evaluating robust foundation models for human genomics},
journal={Nature Methods},
year={2025},
volume={22},
number={2},
issn={1548-7105},
doi={10.1038/s41592-024-02523-z},
}

@inproceedings{NIPSworkshop2025,
title={Open-weight genome language model safeguards: Assessing robustness via adversarial fine-tuning},
author={James R. M. Black and Moritz S. Hanke and Aaron Maiwald and Tina Hernandez-Boussard and Oliver M Crook and Jassi Pannu},
booktitle={NeurIPS 2025 Workshop on Biosecurity Safeguards for Generative AI},
year={2025},
url={https://openreview.net/forum?id=bKLp7oaU1L}
}

@INPROCEEDINGS{oakland2017,
  author={Shokri, Reza and Stronati, Marco and Song, Congzheng and Shmatikov, Vitaly},
  booktitle={2017 IEEE Symposium on Security and Privacy (SP)}, 
  title={Membership Inference Attacks Against Machine Learning Models}, 
  year={2017},
  doi={10.1109/SP.2017.41}
}

@INPROCEEDINGS{oakland2022,
  author={Carlini, Nicholas and Chien, Steve and Nasr, Milad and Song, Shuang and Terzis, Andreas and Tramèr, Florian},
  booktitle={2022 IEEE Symposium on Security and Privacy (SP)}, 
  title={Membership Inference Attacks From First Principles}, 
  year={2022},
  doi={10.1109/SP46214.2022.9833649}
}

@article{science2024,
author = {Eric Nguyen  and Michael Poli  and Matthew G. Durrant  and Brian Kang  and Dhruva Katrekar  and David B. Li  and Liam J. Bartie  and Armin W. Thomas  and Samuel H. King  and Garyk Brixi  and Jeremy Sullivan  and Madelena Y. Ng  and Ashley Lewis  and Aaron Lou  and Stefano Ermon  and Stephen A. Baccus  and Tina Hernandez-Boussard  and Christopher Ré  and Patrick D. Hsu  and Brian L. Hie },
title = {Sequence modeling and design from molecular to genome scale with Evo},
journal = {Science},
volume = {386},
number = {6723},
year = {2024},
doi = {10.1126/science.ado9336},
}

@inproceedings {usenix2019,
author = {Nicholas Carlini and Chang Liu and {\'U}lfar Erlingsson and Jernej Kos and Dawn Song},
title = {The Secret Sharer: Evaluating and Testing Unintended Memorization in Neural Networks},
booktitle = {28th USENIX Security Symposium (USENIX Security 19)},
year = {2019},
isbn = {978-1-939133-06-9},
publisher = {USENIX Association},
}

@inproceedings {usenix2021,
author = {Nicholas Carlini and Florian Tram{\`e}r and Eric Wallace and Matthew Jagielski and Ariel Herbert-Voss and Katherine Lee and Adam Roberts and Tom Brown and Dawn Song and {\'U}lfar Erlingsson and Alina Oprea and Colin Raffel},
title = {Extracting Training Data from Large Language Models},
booktitle = {30th USENIX Security Symposium (USENIX Security 21)},
year = {2021},
isbn = {978-1-939133-24-3},
publisher = {USENIX Association},
}

@inproceedings{usenix2023,
author = {Nicolas Carlini and Jamie Hayes and Milad Nasr and Matthew Jagielski and Vikash Sehwag and Florian Tram{\`e}r and Borja Balle and Daphne Ippolito and Eric Wallace},
title = {Extracting Training Data from Diffusion Models},
booktitle = {32nd USENIX Security Symposium (USENIX Security 23)},
year = {2023},
isbn = {978-1-939133-37-3},
publisher = {USENIX Association},
}

@misc{arc2025opengenome2,
  author = {{Arc Institute}},
  title = {{OpenGenome2}},
  year = {2025},
  howpublished = {Hugging Face dataset},
  url = {https://huggingface.co/datasets/arcinstitute/opengenome2},
  note = {Accessed: 2026-10-02}
}

@misc{minimap2manual,
  author = {Heng Li},
  title = {Minimap2 Manual},
  howpublished = {\url{https://lh3.github.io/minimap2/minimap2.html}},
  note = {Accessed: Oct. 6, 2026}
}

\end{document}